\documentclass{ws-procs9x6}

\begin{document}

\title{
Walls and vortices in supersymmetric \\
non-abelian gauge theories 
}

\author{Youichi~Isozumi\footnote{ 
\uppercase{S}upported by a 21st 
\uppercase{C}entury \uppercase{COE} 
\uppercase{P}rogram at 
\uppercase{T}okyo \uppercase{T}ech "\uppercase{N}anometer-
\uppercase{S}cale \uppercase{Q}uantum \uppercase{P}hysics" by the 
\uppercase{M}inistry of \uppercase{E}ducation, 
\uppercase{C}ulture, \uppercase{S}ports, \uppercase{S}cience 
and \uppercase{T}echnology. 
}
, 
Muneto~Nitta\footnote{\uppercase{S}upported by 
\uppercase{J}apan \uppercase{S}ociety 
for the \uppercase{P}romotion of \uppercase{S}cience 
under the \uppercase{P}ost-doctoral \uppercase{R}esearch 
\uppercase{P}rogram. 
}
, 
 Keisuke~Ohashi\footnote{\uppercase{S}upported by 
 \uppercase{J}apan \uppercase{S}ociety 
for the \uppercase{P}romotion of \uppercase{S}cience 
under the \uppercase{P}ost-doctoral \uppercase{R}esearch 
\uppercase{P}rogram. 
}
, 
 and 
Norisuke~Sakai
\footnote{\uppercase{S}peaker at the conference. 
\uppercase{W}ork partially
supported by \uppercase{G}rant-in-\uppercase{A}id 
for \uppercase{S}cientific \uppercase{R}esearch 
from the \uppercase{M}inistry of \uppercase{E}ducation, 
\uppercase{C}ulture, \uppercase{S}ports, 
\uppercase{S}cience and \uppercase{T}echnology, 
\uppercase{J}apan \uppercase{N}o.13640269 
and 16028203 for the priority area ``origin of mass''. 
}}

\address{Department of Physics, Tokyo Institute of 
Technology \\
Tokyo 152-8551, JAPAN  
}

\maketitle

\abstracts{
We review recent results 
(hep-th/0405194, hep-th/0405129, and hep-th/0404198) 
on the BPS multi-wall solutions in 
supersymmetric 
$U(N_{\rm C})$ gauge theories in five dimensions 
with 
$N_{\rm F}(>N_{\rm C})$ hypermultiplets in the 
fundamental representation. 
Total moduli space of the BPS non-Abelian walls 
is found to be the complex Grassmann manifold 
$SU(N_{\rm F}) / 
[SU(N_{\rm C})\times SU(N_{\rm F}-{N}_{\rm C}) \times U(1)]$. 
Exact solutions are obtained with full generic moduli 
for infinite gauge coupling. 
A $1/4$ BPS equation is also solved, giving 
vortices together with the non-Abelian 
walls and monopoles in the Higgs phase attached to the 
vortices. 
The full moduli space of the $1/4$ BPS solutions 
is found to be holomorphic 
maps from a complex plane to the wall moduli space. 
}

\section{Introduction}

Supersymmetry (SUSY) is useful to 
obtain various branes (solitons) 
for model building in the brane-world 
scenario\cite{HoravaWitten,LED,RandallSundrum}. 
The partial preservation of 
SUSY gives BPS states which are 
solutions of equations of 
motion\cite{WittenOlive}. 
Moreover, the resulting theory 
tends to produce an $ N=1$ 
SUSY theory on the world volume, 
which can provide realistic unified models with the 
desirable properties\cite{DGSW}. 
SUSY also helps to obtain stability of the 
soliton. 
The simplest soliton for the brane-world is the domain wall, 
which should be considered in five dimensions. 
Recently the localized 
massless gauge bosons on a wall has 
been obtained 
using SUSY QED interacting with hypermultiplets 
and tensor multiplets\cite{IOS1,IOS2}. 
We anticipate that walls in non-Abelian gauge theories 
will help to obtain localized non-Abelian gauge bosons. 
These walls are called non-Abelian walls, and are 
interesting in its own right. 

We review our papers 
on various BPS solutions of 
the SUSY $U(N_{\rm C})$ 
gauge theory with $N_{\rm F}(>N_{\rm C})$ flavors of 
hypermultiplets in the fundamental representation 
in spacetime dimensions of five or 
less\cite{INOS1,INOS2,INOS3,Isozumi:2004hr}. 
To obtain discrete vacua, we consider non-degenerate 
masses $m_A$ for hypermultiplets $H^{irA}$, 
and the Fayet-Iliopoulos (FI) parameter is introduced\cite{ANS}. 
We have obtained BPS multi-wall solutions as $1/2$ BPS 
solutions, and various combinations of walls, vortices 
and monopoles in the Higgs phase as $1/4$ BPS solutions. 
By taking the limit of infinite gauge coupling, 
we have obtained exact BPS multi-wall solutions with generic 
moduli parameters covering the complete moduli space 
of walls\cite{INOS1}. 
We found that the moduli space of $1/2$ BPS domain walls 
is given by 
a compact complex manifold, the Grassmann manifold\cite{INOS1} 
$G_{N_{\rm F},N_{\rm C}}\equiv {SU(N_{\rm F}) \over 
SU(N_{\rm C})\times SU(N_{\rm F}-N_{\rm C}) 
\times U(1)}$. 
One should note that this is 
the total moduli space of the multi-wall solutions 
including all the topological sectors, 
and that configurations with smaller number of domain 
walls appear as boundaries of the moduli space. 
We have found that the coexistence of mutually 
orthogonal vortex and the wall can be realized as a $1/4$ 
BPS configuration\cite{INOS2}. 
We also found that this $1/4$ BPS equation admits 
monopoles in the Higgs phase which were found 
recently\cite{mono-Higgs}. 
We have obtained the exact 
solitons in the limit of infinite gauge coupling. 
We also identified  the moduli space to be 
all the holomorphic maps from the complex plane to 
the wall moduli space\cite{INOS2}, 
the Grassmann manifold 
$G_{N_{\rm F},N_{\rm C}}$.

\section{Vacua and BPS Equations for Non-Abelian Walls 
}
\label{sc:model-vacua-BPSeq}

Discrete vacua are needed for walls 
and can be realized by 
mass terms for hypermultiplets 
and $U(N_{\rm C})$ gauge group with the $U(1)$ 
factor group allowing the Fayet-Iliopoulos 
term\cite{ANS}. 
We consider a five-dimensional SUSY model with minimal 
kinetic terms for vector and hypermultiplets whose 
physical bosonic fields are ($W_M, \Sigma$) 
and $H^{i}$, respectively. 
We denote space-time indices by 
$M,N, \cdots=0,1,2,3,4$ with the  metric 
$\eta_{MN}={\rm diag}(+1,-1,-1,-1,-1)$. 
For simplicity, we take the same gauge coupling $g$ for 
$U(1)$ and $SU(N_{\rm C})$ factors. 
The $N_{\rm F}$ flavors of hypermultiplets in the 
fundamental representation are 
combined into an $N_{\rm C}\times N_{\rm F}$ matrix. 
We consider $N_{\rm F}>N_{\rm C}$ to obtain 
disconnected SUSY vacua\cite{ANS} appropriate for constructing walls. 
Our model now has only a few parameters: 
the gauge coupling constant $g$, 
the FI parameter $c>0$ for the $U(1)$ gauge group, 
and  the mass matrix of hypermultiplets  
$(M)^A{}_B\equiv m_A\delta ^A{}_B$ $m_A$. 
We assume  non-degenerate mass parameters 
with the ordering $m_A > m_{A+1}$ for all $A$. 
Then the flavor symmetry reduces to 
$ G_{\rm F} = U(1)_{\rm F}^{N_{\rm F}-1}$. 
After eliminating auxiliary fields, we obtain 
the bosonic part of the 
Lagrangian and the scalar potential 
\begin{eqnarray}
{L}_{\rm bosonic} 
&=& 
-\frac{1}{2g^2}{\rm Tr}\left[ F_{MN}F^{MN}\right]
+\frac{1}{g^2}{\rm Tr}
\left[{
D}_M \Sigma {
D}^M \Sigma \right]
\nonumber \\
&&
+{\rm Tr}
\left[ {
D}^M H^i ({
D}_M H^i)^\dagger \right] -V,
\label{fundamental-Lag2}
\end{eqnarray}
\begin{eqnarray}
V&=& 
\frac{g^2}{4}
{\rm Tr}
\Big[
\left(
H^{1}  H^{1\dagger}  - H^{2} H^{2\dagger} 
- c\mathbf{1}_{N_{\rm C}}
\right)^2 +
 4 H^2H^{1\dagger} H^1H^{2\dagger}
\Big] 
\nonumber\\
&
&
+{\rm Tr}\left[
 (\Sigma H^i - H^i M) 
 (\Sigma H^i - H^i M)^\dagger 
 \right],
\end{eqnarray}
The covariant derivatives are defined as 
$D_M \Sigma = \partial_M \Sigma + i[ W_M , \Sigma ]$, 
$D_M H^{i}=(\partial_M + iW_M)H^{i} 
$, 
and field strength is defined as 
$F_{MN}=\frac{1}{i}[D_M , D_N]
=\partial_M W_N -\partial_N W_M + i[W_M, W_N]$. 

SUSY vacua are realized at vanishing vacuum 
energy, which requires 
both contributions from vector and hypermultiplets 
to vanish. 
Conditions of vanishing contribution from vector 
multiplet read 
\begin{eqnarray}
H^{1}  H^{1\dagger}  - H^{2} H^{2\dagger} 
=c\mathbf{1}_{N_{\rm C}},
\quad 
 H^2 H^{1\dagger}= {0}.
\label{D-term-cond}
\end{eqnarray}
The vanishing contribution to vacuum energy 
from hypermultiplets 
gives the SUSY condition for hypermultiplets as 
\begin{eqnarray}
\Sigma H^i - H^i M
=0 , \  
\label{F-term-cond++}
\end{eqnarray}
for each index $A$.
Non-degenerate masses for hypermultiplets 
dictate that only one flavor $A=A_r$ 
can be non-vanishing 
for each color component $r$ of hypermultiplet scalars 
$H^{irA}$ 
with 
\begin{eqnarray}
 H^{1rA}=\sqrt{c}\,\delta ^{A_r}{}_A,\quad H^{2rA}=0. 
 \label{eq:hyper-vacuum}
\end{eqnarray}
This is called the color-flavor locking vacuum. 
The vector multiplet scalars $\Sigma$ are 
determined 
as 
\begin{eqnarray}
\Sigma ={\rm diag.}(m_{A_1},\,m_{A_2},\,\cdots,\,
m_{A_{N_{\rm C}}}).
\end{eqnarray}
We denote a SUSY vacuum specified by a set of 
non-vanishing hypermultiplet 
scalars with the flavor $\{A_r\}$ for each color 
component $r$ as 
$\langle A_1\,A_2\,\cdots\,A_{N_{\rm C}}\rangle$. 
We usually take $A_1<A_2<\cdots<A_{N_{\rm C}}$. 

Walls interpolate between two vacua at $y=\infty$ 
and $y=-\infty$. 
These boundary conditions at $y=\pm \infty$ 
define topological sectors. 
To obtain domain walls, we assume that 
all fields depend only on coordinate of one extra 
dimension $x^4\equiv y$ and the Poincar\'e invariance 
on the four-dimensional world volume of the wall. 
It implies $F_{MN}(W) =0$, $W_\mu =0$, 
where $x^\mu=(x^0, x^1,x^2,x^3)$ are 
four-dimensional world-volume coordinates. 
Note that $W_y$ need not vanish. 
The Bogomol'nyi completion of the energy density 
of our system can be performed as 
\begin{eqnarray}
{E} 
&=& {1 \over g^2}{\rm Tr}\left({ D}_y \Sigma -
{g^2\over 2}\left(c{\bf 1}_{N_{\rm C}}-H^1H^1{}^\dagger 
+H^2H^2{}^\dagger \right)\right)^2
+
{g^2}{\rm Tr}
\Big[
H^2H^{1\dagger} H^1H^{2\dagger} 
\Big] 
\nonumber \\
&&{}
+ {\rm Tr}\left[
({D}_y H^1 + \Sigma H^1 -H^1M) 
({D}_y H^1 + \Sigma H^1 -H^1M)^\dagger\right] 
\nonumber \\ && {}
+ {\rm Tr}\left[
({D}_y H^2 - \Sigma H^2 +H^2M) 
({D}_y H^2 - \Sigma H^2 +H^2M)^\dagger\right]  
\nonumber \\ && {}
+ c \partial_y{\rm Tr}\Sigma 
- \partial_y \left\{{\rm Tr}
\left[ 
\left(\Sigma H^1 - H^1 M\right)H^1{}^\dagger
+ \left(-\Sigma H^2 
+H^2 M\right)H^2{}^\dagger\right]\right\}. 
\label{eq:bogomolnyi}
\end{eqnarray}
Therefore we obtain a lower bound for the energy of the 
configuration by saturating the complete squares. 
The saturation condition gives the BPS equations 
\begin{eqnarray}
D_y \Sigma = 
{g^2\over 2}\left(c{\bf 1}_{N_{\rm C}}-H^1H^1{}^\dagger 
+H^2H^2{}^\dagger \right), 
\quad 
0
=
g^2 H^1H^2{}^\dagger, 
\label{BPSeq-Sigma}
\end{eqnarray}
\begin{eqnarray}
D_y H^1 
=
-\Sigma H^1 + H^1 M,\qquad 
D_y H^2 = \Sigma H^2 -H^2 M. 
\label{BPSeq-H}
\end{eqnarray}

Let us consider a configuration approaching to 
a SUSY vacuum labeled by 
$\langle A_1A_2\cdots A_{N_{\rm C}}\rangle $ 
at the boundary 
of positive infinity $y=+\infty$, and to 
a vacuum 
$\langle B_1B_2\cdots B_{N_{\rm C}}\rangle $ 
at the boundary 
of negative infinity $y=-\infty$. 
Therefore the minimum energy is achieved by 
the configuration satisfying the BPS 
Eqs.~(\ref{BPSeq-Sigma})-(\ref{BPSeq-H}), 
and the energy for the BPS saturated configuration 
is given by 
\begin{eqnarray}
T_{\rm w}&=&\int^{+\infty}_{-\infty}
\hspace{-1.5em}dy \; {E}
=c 
\left[{\rm Tr}\Sigma \right]^{+\infty}_{-\infty}
=c \left(\sum_{k=1}^{N_{\rm C}}m_{A_k}
-\sum_{k=1}^{N_{\rm C}}m_{B_k}\right), 
\label{eq:tension}
\end{eqnarray}
the second term of the last line of Eq.~(\ref{eq:bogomolnyi}) 
does not contribute, 
since the vacuum condition $\Sigma H^i -H^iM =0$ is satisfied 
at $y=\pm \infty$. 
The BPS equations are equivalent to the preservation of 
$1/2$ SUSY\cite{INOS1}.

\section{BPS Multi-Walls}
\label{BPSWS}

With our sign choice of the FI parameter 
$c >0$, 
$H^2$ vanishes in any SUSY vacuum. 
The BPS equation for $H^2$ gives\cite{INOS3} 
$H^2=0$, and $H^2_0=0$. 
By defining 
a complex $N_{\rm C}\times N_{\rm C}$ invertible matrix function 
$S(y)$ 
\begin{eqnarray}
\Sigma + iW_y \equiv S^{-1}\partial_y S,
\label{def-S}
\end{eqnarray}
we can solve the hypermultiplet BPS equation 
in terms of a $N_{\rm C}\times N_{\rm F}$ constant 
complex matrix $H_0
$ as 
integration constants, which we call moduli matrices: 
\begin{eqnarray}
H^1=S^{-1}H_0^1 e^{My}
.
\label{sol-H}
\end{eqnarray}
Two different sets ($S, H_0
$) and 
$(S', H_0
{}'
)$ 
give the same original fields $\Sigma ,\,W_y, H^i$, 
if they are related by 
$V\in GL (N_{\rm C},{\bf C})$ 
\begin{eqnarray}
 S
&\rightarrow& S' = VS,\quad 
H_0 ^1
\rightarrow
H_0^1{}'=VH_0^1
.
\label{art-sym}
\end{eqnarray} 
This transformation $V$ defines an equivalence class 
among sets of the matrix function and moduli matrix 
$(S, H_0)$ which represent physically 
equivalent results. 
This symmetry  comes from 
the $N_{\rm C}^2$ integration constants in solving 
(\ref{def-S}), and represents 
the redundancy of describing the wall solution 
in terms of $(S, H_0
)$. 
We call this `world-volume symmetry', 

The gauge transformations on the original 
fields $\Sigma ,\,W_y,\,H^1, 
$ 
($
H^1  
\rightarrow 
H^1{}' = U H^1$, 
$\Sigma +iW_y 
\rightarrow 
\Sigma' +iW_y' 
= U\left(\Sigma +iW_y\right)U^\dagger  + U\partial_y U^\dagger $) 
can be obtained by multiplying a unitary matrix $U^\dagger$ 
$
 S\,\rightarrow \,S'=SU^\dagger$, $U^\dagger U=1$, 
without causing any transformations on the moduli 
matrices $H_0$. 
Thus we define 
$
 \Omega \equiv SS^\dagger $, 
which is invariant under the gauge transformations $U$ 
of the fundamental theory. 
Together with 
the gauge invariant moduli matrix $H_0$, 
the BPS equations (\ref{BPSeq-Sigma}) for vector 
multiplets can be rewritten in the following 
gauge invariant form 
\begin{eqnarray}
\partial_y^2 \Omega -\partial_y \Omega 
\Omega^{-1} \partial_y \Omega = g^2 
\left(c\, \Omega - H_0 \,e^{2My} H_0{}^\dagger 
\right). 
\label{diff-eq-S}
\end{eqnarray}
We can calculate uniquely the 
$N_{\rm C}\times N_{\rm C}$ complex matrix $S$ 
from the $N_{\rm C}\times N_{\rm C}$ Hermitian 
matrix $\Omega $ with a 
suitable gauge choice. 
Then all the quantities, $\Sigma ,\,W_y,\,H^1$ and $H^2$ 
are obtained by Eqs.~(\ref{def-S}) and (\ref{sol-H}). 

Since we are going to impose two boundary conditions 
at $y=\infty$ and at $y=-\infty$ to the second order 
differential equation (\ref{diff-eq-S}), the number of 
necessary boundary conditions precisely matches to 
obtain the unique solution. 
Therefore there should be no more moduli parameters 
in addition to the moduli 
matrix $H_0$. 
In the limit of infinite gauge coupling, we 
find explicitly that there are no additional moduli. 
We have also analyzed in detail 
the almost analogous nonlinear 
differential equation in the case of the Abelian gauge 
theory at finite gauge coupling and find no additional 
moduli\cite{IOS1}. 
Thus we believe that we should consider only the moduli 
contained in the moduli matrix $H_0$, 
in order to discuss the moduli space of domain walls.

\section{Moduli Space and Exact Solution for Non-Abelian Walls} \label{MSFNAW}

All possible solutions of parallel domain walls 
in the $U(N_{\rm C})$ SUSY gauge theory with 
$N_{\rm F}$ hypermultiplets 
can be constructed once 
the moduli matrix $H_0$ is given. 
The moduli matrix $H_0$ has a redundancy expressed as 
the world-volume symmetry (\ref{art-sym}) : 
$H_0 \sim V H_0$ with 
$V \in GL(N_{\rm C},{\bf C})$. 
We thus find that the moduli space 
denoted by ${M}_{N_{\rm F},N_{\rm C}}$
is homeomorphic to 
the complex Grassmann manifold 
($\tilde N_{\rm C} \equiv N_{\rm F} -N_{\rm C}$):
\begin{eqnarray}
 {M}_{N_{\rm F},N_{\rm C}}
& 
\simeq 
&
\{H_0 | H_0 \sim V H_0, V \in GL(N_{\rm C},{\bf C})\} 
\nonumber \\
& 
\simeq 
&
 G_{N_{\rm F},N_{\rm C}} 
 \simeq 
{SU(N_{\rm F}) \over 
 SU(N_{\rm C}) \times SU(\tilde N_{\rm C}) \times U(1)}\,.
  \label{Gr}
\end{eqnarray}
This is a {\it compact} (closed) set. 
On the other hand, 
scattering of two Abelian walls is described by 
a nonlinear sigma model 
on a {\it non-compact} moduli space\cite{To,To2,IOS1}. 
This fact 
can be consistently understood, if 
we note that the moduli space 
${M}_{N_{\rm F},N_{\rm C}}$ 
includes all BPS topological sectors: 
\begin{eqnarray}
 {
M}_{N_{\rm F},N_{\rm C}} 
 = \sum_{k=0}^{N_{\rm C}\tilde N_{\rm C}} 
{
M}_{N_{\rm F},N_{\rm C}}^k 
 = {
M}_{N_{\rm F},N_{\rm C}}^0 
  \oplus {
M}_{N_{\rm F},N_{\rm C}}^1 \oplus \cdots 
  \oplus {
M}_{N_{\rm F},
N_{\rm C}}^{N_{\rm C}\tilde N_{\rm C}} ,
 \label{decom}
\end{eqnarray}
where ${
M}_{N_{\rm F},N_{\rm C}}^k$ is 
the moduli space of $k$-walls. 
Consider a $k$-wall solution  
and imagine a situation such 
that one of the outer-most walls 
goes to spatial infinity. 
We will obtain a ($k-1$)-wall 
configuration in this limit. 
This implies that the $k$-wall sector 
in the moduli space is an open set compactified 
by the moduli space of ($k-1$)-wall sectors on its boundary. 
Continuing this procedure 
we will obtain a single wall configuration. 
Pulling it out to infinity we obtain a vacuum state in the end. 
A vacuum corresponds to a point as 
a boundary of a single wall sector 
in the moduli space.
Summing up all sectors,
we thus obtain the total moduli space 
${M}_{N_{\rm F},N_{\rm C}}$ 
as a compact manifold.  



The BPS equation (\ref{diff-eq-S}) for the gauge invariant $\Omega$ 
reduces to an algebraic equation 
in the strong gauge coupling limit, given by 
\begin{eqnarray}
 \Omega_{g \to \infty} 
 = (SS^\dagger)_{g \to \infty}  
 = c^{-1}H_0 e^{2My}H_0^{\dagger}. 
  \label{SS-H0}
\end{eqnarray}
Qualitative behavior of walls 
for finite gauge couplings 
is not so different from that
in  infinite gauge couplings. 
In fact 
we have constructed exact wall solutions 
for finite gauge couplings\cite{IOS1} 
and found that their qualitative behavior 
is the same as the infinite gauge coupling cases 
found in the literature\cite{To,AT
}. 

SUSY gauge theories reduce to 
nonlinear sigma models in general 
in the strong gauge coupling limit 
$g \to \infty$.
This is the HK nonlinear sigma model 
on the cotangent bundle over 
the complex Grassmann manifold\cite{LR,ANS} 
\begin{eqnarray}
\label{T*Gr}
 {M}^{M=0}_{\rm vac} \simeq
 T^* G_{N_{\rm F},N_{\rm C}} \simeq 
 T^* \left[SU(N_{\rm F}) \over 
 SU(N_{\rm C}) \times SU(\tilde N_{\rm C}) \times U(1) 
 \right] \, . 
\end{eqnarray}
From the target manifold (\ref{T*Gr}) one can easily 
see that there exists a duality 
between theories with the same flavor and 
two different gauge groups 
in the case of the infinite gauge coupling\cite{AP,ANS}:
$ U(N_{\rm C}) \leftrightarrow U(\tilde N_{\rm C})$ 
with $\tilde N = N_{\rm F}-N_{\rm C}$. 
This duality holds for the Lagrangian of the 
nonlinear sigma models, and 
leads to the duality of the BPS equations 
for these two theories. 
This duality holds also for the moduli space 
of domain wall configurations.

We can obtain the effective action on the world volume 
of walls, by promoting the moduli parameters to fields 
on the world volume~\cite{Ma}. 
In the case of infinite gauge coupling, the K\"ahler potential 
of the effective Lagrangian is given by~\cite{INOS3} 
\begin{eqnarray}
 {
L}_{\rm walls}
  = c \int d^4 \theta \int dy 
\log \det \Omega
  = c \int d^4 \theta \int dy 
\log \det (H_0 e^{2 M y} H_0{}^\dagger). 
 \label{moduli_metric}
\end{eqnarray}

\section{Meaning of the Moduli Matrix $H_0$ and 
Vortices}
\label{sc:moduli-matrix}

\begin{figure}
\begin{center}
\includegraphics
[width=6cm,height=2.5cm 
]
{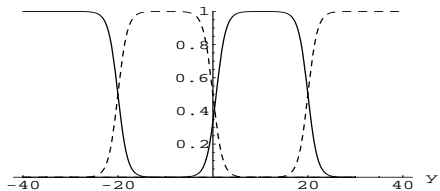}
\caption{
Region of rapid change of hypermultiplets indicates the 
positions of walls. }
\label{fig:wall-hyper}
\end{center}
\end{figure}

To illustrate the physical meaning of the moduli matrix, 
we first consider the Abelian gauge group $N_{\rm C}=1$. 
The moduli matrix in this case can be parametrized as 
\begin{equation}
H_0=\sqrt{c}(e^{r_1}, e^{r_2}, \cdots, e^{r_{N_F}}), 
\label{eq:moduli-abelian}
\end{equation}
where $e^{r_A}$ are complex moduli parameters. 
Let us note that the first entry can be fixed to 
$e^{r_1}=1$ by using the world-volume symmetry 
(\ref{art-sym}). 
Then we obtain the hypermultiplet scalars as 
\begin{equation}
H=S^{-1}H_0 e^{My}=S^{-1} 
\sqrt{c}(e^{r_1+m_1y}, \cdots, e^{r_{N_F}+m_{N_F}y}). 
\end{equation}
The $y$ dependence due to the masses of 
the hypermultiplet flavors shows that 
the relative magnitude 
of different flavors varies as $y$ varies, 
indicating that the solution approaches 
various vacua at various $y$. 
The transition between two adjacent vacua 
occurs when two different flavors 
becomes comparable to each other. 
This transition region is precisely where 
the wall separating two vacua is located. 
The wall location separating $A$- and $A+1$-th 
vacua is then determined in terms of  
the relative magnitude of different flavors of 
the moduli matrix elements 
as ${\rm Re}r_{A}+m_{A}y \sim {\rm Re}r_{A+1}+m_{A+1}y$. 
The overall normalization is taken care of by the function 
$S^{-1}(y)$. 
Consequently the hypermultiplet scalars exhibit the 
multi-wall behavior as illustrated in 
Fig.~\ref{fig:wall-hyper} 
for the case of $N_{\rm F}=4$. 

Similar consideration applies to non-Abelian case. 
The only difference is that some of the matrix elements 
of the moduli matrix $H_0$ can be made to vanish 
by means of the world-volume symmetry 
(\ref{art-sym}). 
The topological sector with the maximal number of moduli 
is given by the vanishing left-lower and right-upper 
triangular parts\cite{INOS3}. 
Therefore the dimension of the moduli space is given by 
\begin{eqnarray}
 \dim_{\bf C} {M}_{N_{\rm F},N_{\rm C}} \equiv 
N_{\rm C} N_{\rm F} -N_{\rm C}^2 
=N_{\rm C} \tilde{N}_{\rm C}. \label{DoF-moduli}
\end{eqnarray}

We have also found interesting characteristic behavior of 
the non-Abelian walls. 
Depending on the quantum numbers of the wall, two walls 
can pass through each other, maintaining their identity. 
We call these pair as penetrable walls\cite{INOS3}. 
On the other hand, certain combinations of walls cannot 
penetrate each other, resulting in impenetrable 
walls. 
If two walls are impenetrable, two walls are compressed 
each other when the relative distance moduli becomes 
negative infinity. 
In the case of Abelian gauge theories, only the impenetrable 
walls can occur\cite{IOS1,To,To2}.


If we make the moduli matrix $H_0$ in 
Eq.(\ref{eq:moduli-abelian}) to depend on 
the world volume coordinates, the wall location 
should depend on the position on the world volume. 
Then the walls will be curved in general. 
If we have a zero in $H_0$, it will give us a 
spike-like behavior of the wall, which becomes a 
vortex ending at the wall\cite{GPTT}. 
As expected for vortices, magnetic fields are also 
generated at the same time. 
Similarly, if we allow an exponential dependence on 
the world volume coordinates for the moduli 
matrix elements, the wall can tilt and a magnetic field 
is generated along the tilted wall. 
In fact we have found that the addition of vortex 
perpendicular to the walls can preserve $1/2$ of 
the surviving SUSY on the world volume 
of the wall. 
Consequently the combined configuration preserves 
$1/4$ of the original SUSY. 
We also found that the $1/4$ BPS equation allows 
another BPS object, the monopole in the Higgs 
phase, which was found recently\cite{mono-Higgs}. 

\begin{figure}[htb]
\includegraphics[width=4.3cm]
{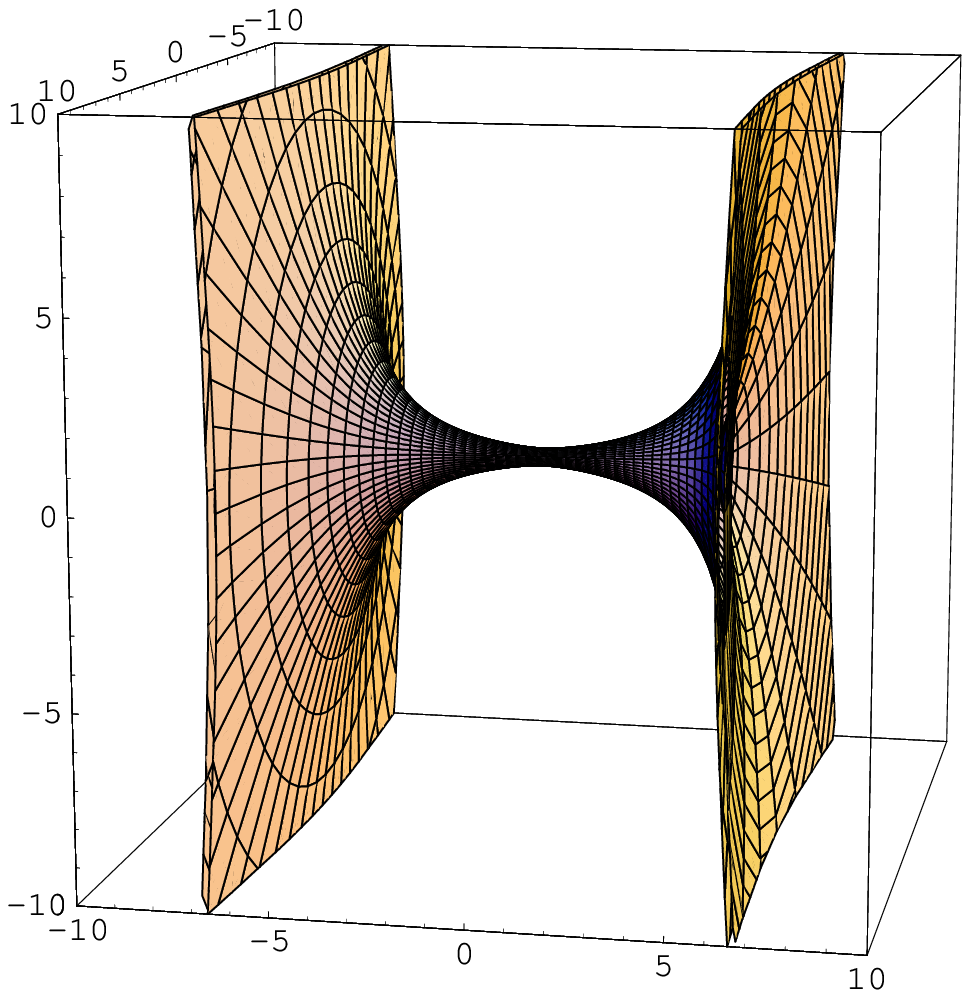}
\put(-75,-2){$x^3$}
\put(-122,60){$x^1$}
\put(-105,115){$x^2$}
\put(-60,-7){a)}
\includegraphics[width=4.3cm]
{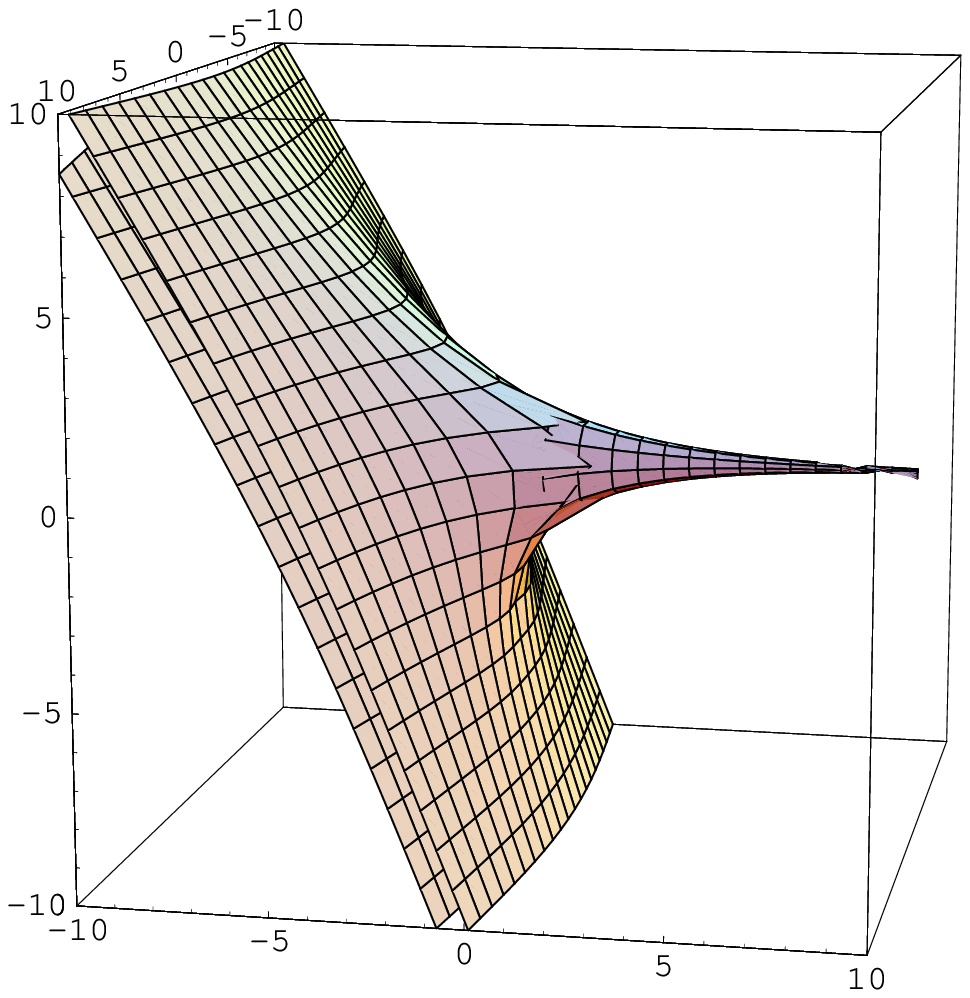}
\put(-75,-2){$x^3$}
\put(-122,60){$x^1$}
\put(-105,115){$x^2$}
\put(-60,-7){b)}
\caption{\label{fig:wormhole} Surfaces defined 
by the same energy density 
: 
a) A vortex stretched between walls with 
$H_0(z)e^{Mx^3}=\sqrt{c}(e^{x^3},ze^{4},e^{-x^3})$. 
b) A vortex 
attached to a tilted wall with 
$H_0(z)e^{Mx^3}=\sqrt{c}(z^2e^{x^3},e^{-1/2z})$.
Note that there are two surfaces with the same energy 
for each wall.}
\end{figure}

We find that the moduli space of solutions of 
this $1/4$ BPS equations is all the holomorphic maps 
from a complex plane to the wall moduli space, 
the deformed complex Grassmann manifold 
$ G_{N_{\rm F},N_{\rm C}}=
{SU(N_{\rm F}) \over 
 SU(N_{\rm C}) \times SU(\tilde N_{\rm C}) \times U(1)}$ 
in Eq.(\ref{Gr}). 
We can obtain exact solutions in the limit of 
infinite gauge coupling\cite{INOS2}. 
As an illustrative example, we show 
a vortex connecting two walls, and 
a vortex ending on a tilted wall in Fig.\ref{fig:wormhole}. 
In the example Fig.\ref{fig:wormhole}a), 
the left-most vacuum outside of the wall 
and the right-most vacuum outside of the wall 
are different. 
They have to touch at the middle of the vortex. 
Therefore there must be a kink separating these 
two vacua. 
This is precisely analogous to the kink in the middle of 
the vortex, where a monopole in the Higgs phase 
resides\cite{mono-Higgs}. 
The example in Fig.\ref{fig:wormhole}b) 
gives a model for a noncommutative 
plane, since there is a magnetic flux flowing along 
the wall. 

\vskip 0.5cm

The authors thank Koji Hashimoto and 
David Tong for a useful discussion.

\bibliographystyle{plain}

\end{document}